\documentclass[10pt,conference]{IEEEtran}
\IEEEoverridecommandlockouts
\usepackage{cite}
\usepackage{amsmath,amssymb,amsfonts}
\usepackage{graphicx}
\usepackage{textcomp}
\usepackage{xcolor}

\usepackage{algorithm, algpseudocode, booktabs, multirow, multicol, pgfplots}
\usepackage{caption}
\usepackage{enumitem}
\usepackage[T1]{fontenc}

\makeatletter
\let\NAT@parse\undefined
\makeatother
\usepackage[hidelinks]{hyperref}

\def\BibTeX{{\rm B\kern-.05em{\sc i\kern-.025em b}\kern-.08em
    T\kern-.1667em\lower.7ex\hbox{E}\kern-.125emX}}

\newcommand{\textttt}[1]{\texttt{\small #1}}

\begin{document}

\title{Software Entity Recognition with\\ Noise-Robust Learning}

\author{
  \IEEEauthorblockN{
    Tai Nguyen\IEEEauthorrefmark{3}\IEEEauthorrefmark{1}\IEEEauthorrefmark{5},
    Yifeng Di\IEEEauthorrefmark{2}\IEEEauthorrefmark{1},
    Joohan Lee\IEEEauthorrefmark{4}\IEEEauthorrefmark{5},
    Muhao Chen\IEEEauthorrefmark{4},
    Tianyi Zhang\IEEEauthorrefmark{2}
  }

  \IEEEauthorblockA{
    \IEEEauthorrefmark{2}Purdue University, West Lafayette, USA
  }

  \IEEEauthorblockA{
    \IEEEauthorrefmark{3}University of Pennsylvania, Philadelphia, USA
  }

  \IEEEauthorblockA{
    \IEEEauthorrefmark{4}University of Southern California, Los Angeles, USA
  }
  
  \IEEEauthorblockA{
  taing@seas.upenn.edu, di5@purdue.edu, joohanl@usc.edu, muhaoche@usc.edu, tianyi@purdue.edu
  }
}

\maketitle

\begingroup\renewcommand\thefootnote{\IEEEauthorrefmark{1}}
\footnotetext{Equal contribution.}
\endgroup

\begingroup\renewcommand\thefootnote{\IEEEauthorrefmark{5}}
\footnotetext{Work done as remote research interns at Purdue University.}
\endgroup

\begin{abstract}
Recognizing software entities such as library names from free-form text is essential to enable many software engineering (SE) technologies, such as traceability link recovery, automated documentation, and API recommendation. While many approaches have been proposed to address this problem, they suffer from small entity vocabularies or noisy training data, hindering their ability to recognize software entities mentioned in sophisticated narratives. To address this challenge, we leverage the Wikipedia taxonomy to develop a comprehensive entity lexicon with 79K unique software entities in 12 fine-grained types, as well as a large labeled dataset of over 1.7M sentences. Then, we propose \textit{self-regularization}, a noise-robust learning approach, to the training of our software entity recognition (SER) model by accounting for many dropouts. Results show that models trained with self-regularization outperform both their vanilla counterparts and state-of-the-art approaches on our Wikipedia benchmark and two Stack Overflow benchmarks. We release our models\footnote{\url{https://huggingface.co/taidng/wikiser-bert-base};\\\url{https://huggingface.co/taidng/wikiser-bert-large}.}, data, and code for future research.\footnote{\url{https://github.com/taidnguyen/software_entity_recognition}}
\end{abstract}

\begin{IEEEkeywords}
Software Entity Recognition, Datasets, Noise-Robust Learning
\end{IEEEkeywords}

\section{Introduction}
Software entity recognition (SER) is an integral task for acquiring software-related knowledge. It serves as the backbone of many downstream software engineering applications, such as traceability link recovery~\cite{antoniol2002recovering, marcus2003recovering, bacchelli2010linking, dagenais2012recovering}, automated documentation~\cite{subramanian2014live, treude2016augmenting, chen2016mining, chen2016techland, li2018improving}, API recommendation~\cite{huang2018api, rahman2016rack, xie2020api}, and bug fixing~\cite{chen2015crowd, gao2015fixing, mahajan2020recommending, mahajan2022providing}.

Early work in this research direction employs pattern-matching methods to identify software entities based on pre-defined linguistic patterns or predefined dictionaries~\cite{rigby2013discovering, subramanian2014live, gupta2013part, capobianco2013improving}. However, these methods lack the flexibility to handle the sophistication and ambiguity in free-form text~\cite{huo2022arclin}. Machine learning methods have been increasingly adopted to solve this task~\cite{ye2016software, chen2019sethesaurus, tabassum2020code, zhou2020improving, malik2022software, gorbatovski2023bayesian}. 
For example, S-NER~\cite{ye2016software} uses a feature-based Conditional Random Field (CRF) model to recognize software entities in five categories, including {\em Programming Language}, {\em Platform}, {\em API}, {\em Tool-library-framework} and {\em Software Standard}. However, S-NER is trained on a small dataset with 4,646 sentences and 2,404 named entities. It 
does not generalize well to commonly mentioned entities such as ``AMD64'' and ``memory leak''. Furthermore, given the simplicity of its model design, it only achieves a 78\% F1 score on a Stack Overflow dataset.

In general text domains, deep learning models, such as BiLSTM-CRF~\cite{huang2015bidirectional} and BERT-NER~\cite{devlin2018bert}, have emerged as the current paradigms for Named Entity Recognition (NER). However, these models can only detect entities in general text domains, such as person names and locations. Due to the domain shift challenge, simply finetuning them to a highly specialized domain such as software engineering is not sufficient~\cite{tabassum2020code}. In recent years, an approach called SoftNER~\cite{tabassum2020code} has been proposed to detect fine-grained software entity types with BERTOverflow, a BERT model finetuned on Stack Overflow data. Their evaluation shows it has greatly outperformed BiLSTM-CRF and BERT-base models and can detect various types of software entities, such as operating systems and software libraries.

Despite the great stride, our assessment shows that existing models, including SoftNER, still fall short of addressing domain shift, limited vocabularies, and morphological names in software engineering. In particular, the training data of SoftNER is noisy, since it is constructed synthetically based on Stack Overflow (SO) tags, which can be created by any SO users and suffer from informal naming conventions and all sorts or randomness. Our manual analysis shows that the training data of SoftNER has a high labeling error rate of 17.79\% (detailed in Section~\ref{sec:manual_validation}). We hypothesize that it may be due to the lack of widespread use of double annotation and metadata for automatic annotation. This motivates us to construct a new dataset with fewer labeling errors but more sentences and named entities.

To address this limitation, we develop an automated pipeline to develop a large, high-quality software entity dataset based on Wikipedia. We call this dataset \textsc{WikiSER}. Compared with Stack Overflow, Wikipedia strives to be a comprehensive online encyclopedia. It generally exhibits better structures, well-formedness, grammaticality, and semantic coherence in its natural language sentences~\cite{yano2016taking}. Since Wikipedia contains articles in numerous domains, our approach first processes the Wikpedia taxonomy starting from the root category ``Computing'' and performs hierarchical pruning to only retain SE-related categories. Then, it extracts the titles of all articles belonging to these categories as well as their aliases curated by Wiki authors as the entity lexicon.
In this way, we get overall 79K software entities in 12 fine-grained categories, e.g., algorithms, data structures, libraries, and OS.

Since Wikipedia articles often contain hyperlinks to other articles, each hyperlinked word or phrase can be treated as a mention of another entity, which can be leveraged to curate the text corpus with labeled entities. Based on our observation, Wiki authors typically only add hyperlinks to the first mention of an entity in a Wikipedia article. Thus, we further develop a matching method to automatically propagate entity types, so that we can obtain more sentences with labeled entities. In the end, we curate a large corpus with 1.7M sentences labeled with the 79K entities.
Our manual validation demonstrates that the labeling error rate of our dataset is 9.17\%, compared with an error rate of 17.79\% in the SoftNER dataset (Section~\ref{sec:manual_validation}).

Furthermore, we propose a noise-robust learning framework called {\em self-regularization} that trains an SER model to be consistent with its predictions under a noisy setting. Specifically, to enhance the robustness of model training, our framework leverages the dropout mechanism to simulate the prediction inconsistency from multiple differently initialized models and incorporates an agreement loss as a regularization mechanism, encouraging prediction consistency in the presence of noisy labels. By relying solely on the training data, our framework offers several advantages over other noise-robust methods and can be easily adapted to any model initialization.

Our evaluation demonstrates that BERT models trained with our self-regularization framework outperform multiple baselines. Specifically, our self-regularized BERT$_{base}$ model outperforms a SOTA SER model called SoftNER~\cite{tabassum2020code} by 7.1\% in terms of F1 score. Furthermore, self-regularization also outperforms co-regularization~\cite{zhou2021learning}, a SOTA noise-robust learning method, by 2.9\% in F1. This performance gain from self-regularization also generalizes to two existing SER datasets~\cite{tabassum2020code, ye2016learning} obtained from Stack Overflow. Finally, we observe that self-regularization is more effective for smaller models and in-domain training indeed plays a major role in boosting the performance gain of SER in different types of data, e.g., Wikipedia vs.~Stack Overflow. 

Overall, these findings provide valuable insights into the strengths and limitations of our approach.

To sum up, we make the following contributions:

\begin{enumerate}[leftmargin=1.5em]
    \item {\bf Dataset.} We leverage Wikipedia to develop a comprehensive lexicon of 79K software entities in 12 fine-grained categories, as well as a large labeled dataset with 1.7M sentences and 3.4M entity labels. We make our dataset publicly available to cultivate future research.
    \item {\bf Model.} We propose a new noise-robust learning framework that regularizes the training of SER models via a dropout mechanism to account for labeling errors in SER datasets.
    \item {\bf Evaluation.} We conduct a comprehensive evaluation of the proposed approach against the state-of-the-art SER models on multiple datasets.
\end{enumerate}

\section{Related Works}
\label{section_relatedworks}
\subsection{Software Entity Recognition}
Recognizing software entities from text documents has been a long-standing research problem in Software Engineering~\cite{antoniol2002recovering, marcus2003recovering}. Early approaches rely on keyword or rule-based pattern matching to identify software entities, and they mainly focus on identifying API names~\cite{antoniol2002recovering, bacchelli2010linking, dagenais2012recovering, rigby2013discovering, subramanian2014live, treude2016augmenting, ma2019easy}. For example, Bacchelli et al.~design lightweight regular expressions based on common naming conventions to identify class and function names in email discussions~\cite{bacchelli2010linking}. Rigby et al.~encode regular expressions into an island parser to recognize classes, methods, and fields mentioned in Stack Overflow posts \cite{rigby2013discovering}.

More recently, machine learning, especially deep learning, has been increasingly adopted for software entity recognition~\cite{ye2016software, ye2016learning, zhou2018recognizing, tabassum2020code, zhou2020improving, huo2022arclin}.  These approaches also recognize a richer set of software entities beyond API names. For example, Ye et al.~proposed a Conditional Random Field (CRF) model to identify five types of software entities in Stack Overflow posts, including {\em programming languages}, {\em platforms}, {\em APIs}, {\em software libraries and frameworks}, {\em software standards}~\cite{ye2016software}. They further integrated word embeddings with the CRF model to address the challenges of polysemy and naming variations~\cite{ye2016learning}. 
 Zhou et al.~proposed a similar word embedding-based CRF model but focused on identifying software entities in bug reports \cite{zhou2018recognizing}. More recently, they  proposed a BiLSTM-CRF model for software entity recognition~\cite{zhou2020improving}.  
Chen et al.~proposed to identify morphological relations between software entities by analyzing and comparing the word embeddings learned from software-related documents and general text documents \cite{chen2017unsupervised}. Tabassum et al.~finetuned BERT with Stack Overflow posts and proposed a BERT-CRF model called SoftNER to detect software entities in Stack Overflow posts \cite{tabassum2020code}. Huo et al.~proposed to combine BiLSTM-CRF with a context-aware scoring mechanism to identify API mentions in free-form text \cite{huo2022arclin}. 

Recently, Chew et al.~\cite{chew2022comparative} conducted a comparative evaluation on S-NER~\cite{ye2016software}, Stanford-NER~\cite{finkel2005incorporating}, BERT~\cite{devlin2018bert}, and BERTOverflow~\cite{tabassum2020code}. They found that S-NER achieved the best performance while BERTOverflow achieved the worst performance. However, the best model only achieved 78.18\% F1-score. Our work advances the state-of-the-art by addressing the data noise challenge in SER. To achieve this, we present a large NER dataset with fewer noisy labels and a noise-robust learning framework to account for data noises coming from annotation errors and ambiguous software entity names. Our evaluation shows that NER models trained with our new dataset and noise-robust learning framework achieved the best performance among six NER baselines.

\subsection{Named Entity Recognition}

Our work is also closely related to the general-domain Named Entity Recognition (NER) task in natural language processing (NLP) ~\cite{nadeau2007survey}. NER models aim to identify named entities in the general text domain, such as persons, organizations, and locations. Recently, deep learning models have gained dominance in obtaining state-of-the-art results. These models capture complex interactions between words and their contexts by learning from large text corpora labeled with named entities. Huang et al.~proposed a BiLSTM-CRF model to encode the contextual information in a sentence for NER~\cite{huang2015bidirectional}. Following this work, many BiLSTM-CRF-based models have been proposed~\cite{lample2016neural, chiu2016named, nguyen2016toward, zheng2017joint, zhou2017joint, ma2016end, tran2017named, rei2016attending, wei2016disease, lin2017multi}.

Recently, Transformer-based language models~\cite{vaswani2017attention} have become a new standard for developing NER models. Researchers have experimented with a range of pretrained language models for NER, such as BERT~\cite{devlin2018bert}, RoBERTa~\cite{liu2019roberta}, LUKE~\cite{yamada2020luke} and even autoregressive models such as GPT~\cite{radford2018improving}. Typically, these language models are first pretrained on a large unlabeled text corpus through self-supervised learning and then finetuned for a specific task.

It remains challenging to reuse NER models trained on general text corpora to highly specialized domains such as biology and medicine, as shown by previous studies~\cite{yang2017transfer, jia2019cross, jia2020multi, liu2021crossner}. Several known challenges present, including domain-specific naming standards, common word polysemy~\cite{ye2016learning, islam2018comparison}, and naming variations~\cite{chew2022comparative}. Thus, developers have spent great effort to develop domain-specific NER models, such as BioBERT~\cite{lee2020biobert} for biomedical literature, ClinicalBERT~\cite{alsentzer2019publicly} for clinical documents, or SciBERT~\cite{beltagy2019scibert} for scientific literature. Our work focuses on doing NER for software documents.

\subsection{Noise-robust Learning}

Training data is often noisy and contains various types of errors, which can degrade the performance of ML models. Noise-robust learning has been widely studied in computer vision~\cite{zhang2018generalized, wang2019symmetric, sukhbaatar2014training, goldberger2017training, wang2017multiclass, chang2017active, muller2019does, srivastava2014dropout}. Recently, several approaches have investigated noise-robust learning in NLP  tasks~\cite{wang2019crossweigh, xiao2019quantifying, zhou2021learning, wang2019learning, cheng2020learning}. Wang et al.~\cite{wang2019crossweigh} proposed CrossWeigh, a method that partitions the training data into several folds and trains independent NLP models to identify potential noisy labels. However, this approach requires training multiple models on different data folds and is thus computationally expensive and only supports fold-level noise estimation. Xiao et al.~\cite{xiao2019quantifying} proposed a Bayesian Neural Network (BNN) method that quantifies model and data uncertainties for NER and sentiment analysis tasks. Wang et al.~\cite{wang2019learning} proposed NetAb, presupposing that noise can be simulated by flipping clean labels randomly. However, this presupposition is overturned by Cheng et al.~\cite{cheng2020learning}, indicating that different datasets have different noise rates. Zhou and Chen \cite{zhou2021learning} proposed a co-regularization framework that consists of two or more neural networks with the same structure but different initializations, which is particularly effective at reducing the impact of noise in the training data and improving the accuracy of the NER model. Inspired by this approach, we propose \textit{Self-regularization}, a noisy-robust learning approach for NER in the SE domain. Self-regularization outperforms co-regularization in our evaluation while requiring training of a single model instead of simultaneously many models, making it more computationally efficient.

\section{Problem Formulation}
\label{section_formulation}

This section defines the research problem of recognizing software entities from text documents. 

\textbf{Definition 1. (Software Entity):} Software entities are nouns and noun phrases that describe specific objects, concepts, and procedures related to software engineering, such as an algorithm name and a data structure name. To effectively perform software entity recognition, it is crucial to establish a well-defined and easily interpretable inventory of entity types. For software entities, our primary objective is to construct a domain-specific inventory of entity types that comprehensively cover various aspects of software engineering knowledge. Therefore, we design our inventory of entity types to cover software engineering concepts exclusively. In future work, one can extend our dataset with more software-related entities, such as software engineering conference names and computer scientist names, based on the downstream applications.

With this domain focus, three of the authors conduct an iterative process involving a focus group over three 2-hour sessions. Each author independently annotated 50 samples of software entities from our corpus. Then, they compared notes and reconciled differences through discussions. After three iterations of sampling, annotation, and consensus building, they ultimately reached an agreement to center our attention on the following 12 fine-grained software entity types that cover key software entities while balancing specificity versus coverage. These 12 types are {\em Algorithm}, {\em Application}, {\em Architecture}, {\em Data structure}, {\em Device}, {\em Error name}, {\em General concept}, {\em Language}, {\em Library}, {\em License}, {\em Operating system}, and {\em Protocol}. We provide a definition and examples for each software entity type below.

\begin{itemize}

\item \textbf{Algorithm.} This type includes computational procedures, algorithms, and paradigms that take inputs and perform defined operations to produce outputs, e.g., Bubble Sort, Auction Algorithm, and Collaborative Filtering.

\item \textbf{Application.} This type includes computer software and programs designed to perform specific user-oriented tasks, e.g., Adobe Acrobat, Microsoft Excel, and Zotero.

\item \textbf{Architecture.} This type includes computer architectures and other related computer system designs, e.g., IBM POWER architecture, Skylake (microarchitecture), and Front-side Bus.

\item \textbf{Data structure.} This type includes standardized ways of organizing and accessing data in computer programs, e.g., Array, Hash table, and mXOR linked list.

\item \textbf{Device.} This type includes physical computing components designed for specific functions, e.g., Samsung Gear S2, iPad, and Intel T5300.

\item \textbf{Error name.} This type includes program errors, exceptions, and anomalous behaviors in computer software, e.g., Buffer Overflow, Memory Leak, and Year 2000 Problem.

\item \textbf{General concept.} This type includes a broad range of programming strategies, paradigms, concepts, and design principles, e.g., Memory Management, Adversarial Machine Learning, and Virtualization.

\item \textbf{Language.} This type includes programming languages and domain-specific languages designed to communicate instructions to computers, e.g., C++, Java, Python, and Rust.

\item \textbf{Library.} This type includes software libraries, packages, frameworks, and other types of APIs, e.g., Beautiful Soup, FFmpeg, and FastAPI.

\item \textbf{License.} This type includes legal terms governing the usage and distribution of software, e.g.,  Cryptix General License, GNU General Public License, and MIT License.

\item \textbf{Operating system.} This type includes system software responsible for managing computer hardware and software resources and providing services for computer programs, e.g., Linux, Ubuntu, Red Hat OS, and MorphOS.

\item \textbf{Protocol.} This type includes rules and standards that define communication between electronic devices, e.g., TLS, FTPS, and HTTP.
  
\end{itemize}

\textbf{Definition 2. (Software Entity Recognition):} We formulate the task of software entity recognition as a token-level classification problem. Given {\em T}, a free-form text in the context of software engineering, the software entity recognition task is to identify every span of words $s = <w_1 w_2 \cdots w_n>$ that refers to a software entity from {\em T} and classify each {\em s} into one of the 12 entity types we defined.

\begin{figure}[t]
    \centering
    \includegraphics[width=\columnwidth]{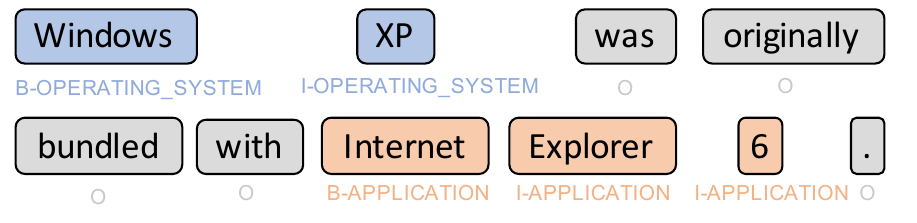}
    \caption{Software entity tagging with the IOB scheme}
    \label{fig_iob}
\end{figure}

In our problem setting, we consider the IOB~\cite{ramshaw1999text} scheme for entity labeling. IOB  is a commonly used tagging format for annotating tokens in NER. It provides a simple way to identify entity boundaries. In the IOB scheme, each token in a sequence is labeled as either {\em B} (i.e., beginning of an entity), {\em I} (i.e., in the middle of an entity), or {\em O} (i.e., not an entity). Figure~\ref{fig_iob} illustrates the IOB labeling scheme with an example sentence from Wikipedia. In this sentence, ``Windows XP'' are labeled as {\em Operating System} and ``Internet Explorer 6'' are labeled as {\em Application}. Since both of them contain multiple words, the first words in them are labeled with {\em B-OPERATION\_SYSTEM} and {\em B-APPLICATION} respectively, while the remaining words are labeled as {\em I-OPERATION\_SYSTEM} and {\em I-APPLICATION}. In the software entity recognition task, an entity is considered correctly recognized {\em only if} the labels of all words in the entity span are correctly predicted.

\section{Dataset Construction}
\label{section_dataset}

\begin{figure*}[t]
    \centering
    \includegraphics[width=2\columnwidth]{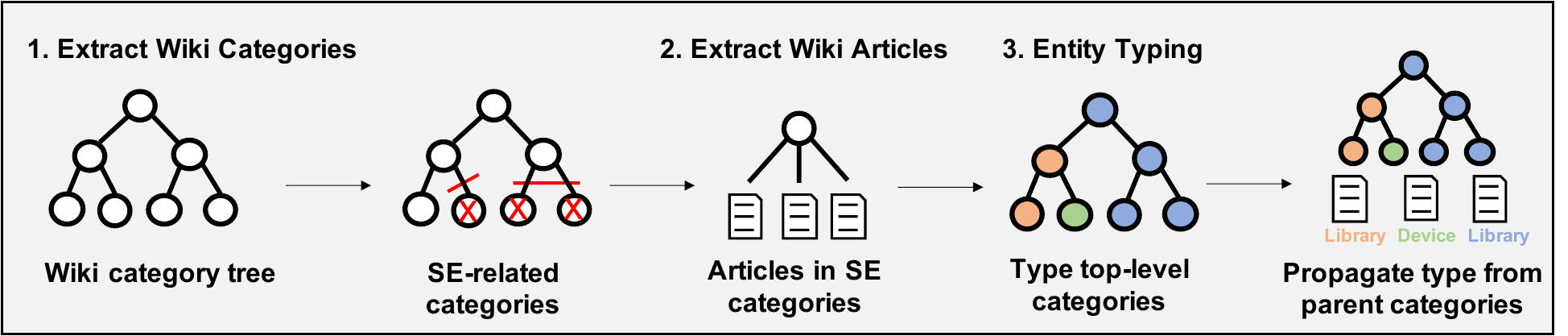}
    \caption{\textsc{WikiSER} Construction Pipeline}
    \label{fig_datapipeline}
\end{figure*}

In this section, we outline the process of identifying and categorizing software entities in Wikipedia corpora. 
Figure~\ref{fig_datapipeline} illustrates the data construction pipeline. The resulting dataset, which we refer to as \textsc{WikiSER}, comprises 1.7M 
sentences labeled with 79K unique software entities (3.4M labels in total).

\subsection{Pruning Wikipedia Taxonomy}
\label{pruning}

Since articles on Wikipedia cover a variety of domains, we need to first prune it to retain ones only related to software engineering. Wikipedia provides a hierarchical classification of its articles based on the domain and topic. In the hierarchy, more general categories appear closer to the root, while more specific categories appear at the bottom. To collect SE categories, we start from \textttt{Category:Computing}\footnote{https://en.wikipedia.org/wiki/Category:Computing}, which is the most general category related to SE. We use the MediaWikiAPI to recursively find all descending subcategories of \textttt{Category: Computing} in the taxonomy (a total of 2M categories).

Note that not all descendants of \textttt{Category: Computing} fall into the 12 SE types we focus on, since our category taxonomy focuses specifically on software entities rather than peripheral topics. For example, \textttt{Category: Computer specialists}\footnote{https://en.wikipedia.org/wiki/Category:Computer\_specialists} is a descendant of \textttt{Category: Computing} but contains experts and researchers in computer science and adjacent fields, which are not our main focus. However, entities labeled as \textttt{Category: Computer specialists} during Wikipedia crawling could be retained in a separate type for future use. Our taxonomical design and ML architecture allow seamless integration of new entity categories if required. To prune the category taxonomy, two of the authors examine the sub-categories of \textttt{Category: Computing} in a top-down manner and create a list of 924 categories that are irrelevant to the 12 software entity types defined in Section~\ref{section_formulation}.
We remove all categories that are descendants of categories in this list. 
Due to Wikipedia's tree structure, removing a parent category translates to a removal of an entire branch. After pruning, we retain 8,469 categories. 

We manually filter the remaining 8,469 categories to ensure their correctness. To facilitate a consensus on this manual filtering, the two authors had a 1-hour discussion regarding the filtering criteria. They also practiced on 50 categories together to reinforce their understanding of the selection criteria. The complete annotation process is divided into three steps. First, each author filters half of the 8,469 categories. They discuss with each other when they are uncertain about a category during the filtering process. Then, they cross-validate their filtering results and discuss the categories where they disagree. The agreement level between the two authors is 0.86 in terms of Cohen's Kappa~\cite{cohen1960coefficient}, indicating a substantial agreement level. Third, they work in pairs to resolve all categories that are flagged as questionable. This manual process took about 62 person-hours. A total of 945 categories are deemed unrelated to the 12 software entities by both authors and 7,524 categories remain after the manual filtering.

\subsection{Collecting Software Entities}
\label{sec:article-collection}


\begin{table}[t]
\caption{Heuristics for Identifying Software Entities in Wikipedia taxonomy}
\resizebox{\linewidth}{!}{
\begin{tabular}{@{}lrr@{}}
\toprule
\multicolumn{1}{c}{Heuristics} & Article \#  & Precision\\ 
\midrule
Contains 1 SE category & 139,752 & 79.3\%    \\ 
{\bf Contains 2 or more SE categories} & 79,899  & \textbf{92.8}\% \\
20\% of labeled categories are SE   & 105,793 & 83.5\%    \\
50\% of labeled categories are SE   & 38,043  & 88.2\%    \\
60\% of labeled categories are SE   & 15,528  & 90.6\%    \\
\bottomrule
\end{tabular}
}
\label{tab_page-criteria}
\end{table}
On Wikipedia, each article is labeled with one or more Wikipedia categories by default. We leverage this categorization to find Wikipedia articles that may describe software entities. Specifically, we write a script to automatically extract Wiki articles labeled with at least one of the 7,524 SE categories identified from the previous step. 139,752 Wiki articles are extracted after this step. 

However, we notice that this corpus is noisy. While many articles are labeled with a SE category, they do not actually describe a specific software entity. For example, ``Software studies'' 
is a Wiki article under the \textttt{Software} category, but it does not refer to a specific software entity. After manually examining 30 articles, we find that those articles are labeled with not only a SE category but also some categories not closely related to SE. For example, the ``Software studies'' article is labeled with \textttt{Computing culture}, \textttt{Cultural studies}, \textttt{Digital humanities}, in addition to \textttt{Software}. 

Based on this insight, we experiment with several heuristics that filter Wikipedia articles by the number or percentage of SE categories among all the labeled categories. Table~\ref{tab_page-criteria} describes each heuristic. We measure the precision of each heuristic. Here, precision refers to how many articles classified as SE-related are actually related to SE. The two authors sample 385 articles from all the extracted Wiki articles. This sample size is considered statistically significant with a 95\% confidence level and a margin of error of 5\%. We then manually label whether the articles are SE-related as the ground truth and compare them with those obtained by filtering based on various heuristics. Consequently, we find that the heuristic of selecting articles labeled with two or more SE categories achieves the highest precision at 92.8\% among all heuristics. The filtered number of articles is 79,899, which has not decreased significantly compared to the number number of articles before filtering.

\subsection{Labeling Software Entity Spans}
As explained in Section~\ref{section_formulation}, we need to identify the span of each software entity mentioned in a Wiki article. We leverage the hyperlinks in a Wiki article, as well as keyword matching, to identify the mentions of software entities in a Wiki article. Specifically, we treat the title of each of the 79,899 articles found in the previous step as a software entity. If a word or a phrase in a sentence of a Wiki article is hyperlinked to a Wiki article in the 79,899 articles, we consider that it mentions a software entity. 

We observe that not all mentions of a software entity are hyperlinked in a Wiki article. Specifically, many Wiki articles only hyperlink the first mention of an entity to its corresponding article. Based on this observation, we further develop a keyword-matching method to identify the mentions of software entities.

A major challenge in this step is that the same entity can be expressed in different forms. For example, ``Long Short-Term Memory'' is often written as ``LSTM''. To address this challenge, we leverage the page redirection mechanism in Wikipedia to recognize aliases, which is commonly adopted in prior work~\cite{torisawa2007exploiting, cucerzan2007large, richman2008mining, nothman2013learning}. In Wikipedia, accessing an article can sometimes automatically send visitors to another article with the same concept but a different name, which is called a redirect. For example, ``LSTM'' is a redirect of ``Long Short-Term Memory''. Hence, when users want to visit the Wikipedia article of ``LSTM'', it will automatically jump to the article of ``Long Short-Term Memory''. Since they both point to the same article, we can safely assume that ``LSTM'' is an alias for ``Long Short-Term Memory''. Using this mechanism, we get aliases for all software entities in our dataset. Given a Wikipedia article, we first perform lemmatization to handle words in different forms and then identify the mentions of a software entity or its alias via exact keyword matching.

\begin{figure}[t]
\vspace{-0.1in}
    \centering
    \includegraphics[width=1\columnwidth]{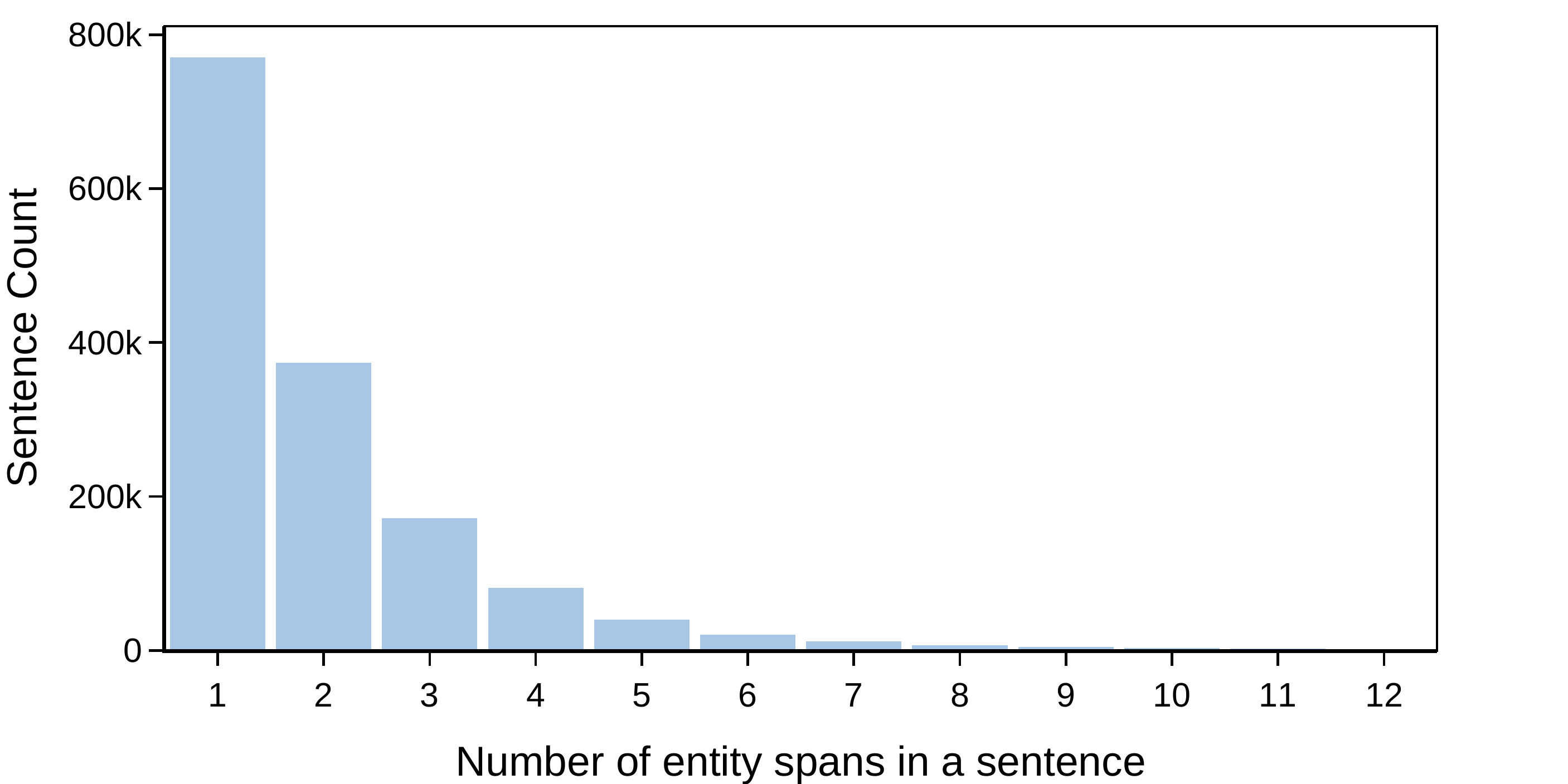}
    \caption{Distribution of Entity Spans per Sentence}
    \label{fig_sentence_entity_distribution}
\end{figure}

In total, we obtain 3.4M sentences from 79,899 articles. Many of these sentences do not mention any software name entities and provide little value for model training and evaluation. Thus, we remove over 1.7M of these instances along with duplicated sentences. Finally, the dataset results in 1,663,431 sentences that mention at least one software entity.  Figure~\ref{fig_sentence_entity_distribution} shows the distribution of the number of name entities in the sentences in \textsc{WikiSer}.

\subsection{Labeling Entity Types}
As the final labeling step, we need to further assign each entity span to the corresponding entity type as defined in Section~\ref{section_formulation}. Note that under the Wikipedia taxonomy, an article belongs to one or more categories. Thus, an intuitive idea is to leverage the categories to infer the type of the entity an article refers to. Based on this idea, we first establish a mapping between the 7,524 categories from Section~\ref{pruning} and the 12 software entity types defined in Section~\ref{section_formulation}. Then, we infer the entity types of an article based on the types of categories they belong to. We elaborate on these two steps below.

{\bf Map Wiki categories to entity types.} To do this, two co-authors first collectively label the SE categories up to the 5th level in the Wikipedia taxonomy, resulting in a total of 1,160 categories. Similar to the manual labeling process in Section~\ref{pruning}, they first discuss the categorization criteria for 1 hour and practice together on 50 categories to enhance consensus on categorization. Then, each of them is assigned half of the categories and is asked to report any disagreement they are uncertain about. The Cohen's Kappa~\cite{cohen1960coefficient} score is 0.82, indicating substantial agreement. They further discuss and resolve any disagreement.
In this way, we manually establish a mapping between the 1,160 categories up to the 5th level to the 12 software entity categories. Then, we developed an automated script that performs a breadth-first traversal of the category hierarchy, starting from categories of level 6, and automatically assigns an entity type to a descending category based on the type of its parent category. Ultimately, we establish a mapping between all 7,524 categories to the 12 entity types.

{\bf Inferring Entity Types.}
Having obtained the inferred types for all SE categories, our goal is to classify the Wikipedia article of each entity based on its categories. However, as explained in Section~\ref{sec:article-collection}, each article can have multiple categories, which may belong to different entity types. For example, consider the software entity ``ChromeOS'', which should be categorized as an {\em Operating System}. However, while it has categories that belong to the {\em Operating System} type, such as \textttt{Category: ARM operating systems} and \textttt{Category: Google operating systems}, it also has a category that belongs to the {\em Application} type (i.e., \textttt{Category: Google Chrome}).

\begin{figure}[t]
    \centering
    \includegraphics[width=\columnwidth]{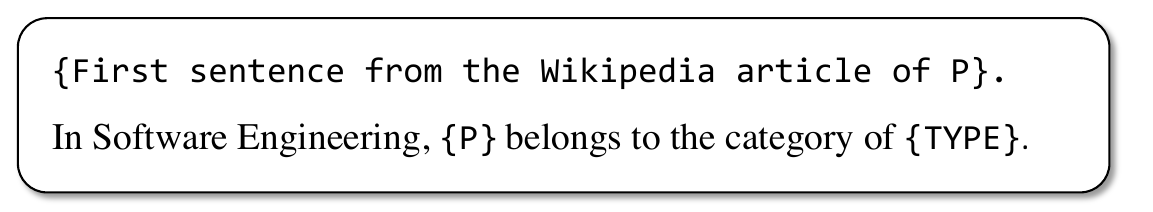}
    \caption{Prompt input to Flan-T5 to infer Entity Type (Content in curly brackets are placeholders)}
    \label{fig:prompt}
\end{figure}

To address this issue, we design three heuristics to decide the final entity type of an article with multiple categories. First, we simply assign the entity type of the most fine-grained category of the article, which is measured by the distance to the root \textttt{Category: Computing}. Second, if all categories of the article are at the same granularity in the hierarchy, we assign the entity type that the majority of categories belong to. Third, if there is still a tie, we infer the entity type of the article by prompting a large language model. Specifically, for a Wikipedia article $P$, we prompt Flan-T5 XL~\cite{chung2022scaling} with the format shown in Figure~\ref{fig:prompt}. Specifically, we prepend the prompt with the first sentence extracted from $P$. We substitute the second mask token with each candidate type and use Flan-T5 XL to calculate the perplexity of each completed prompt. Perplexity measures the degree of uncertainty of the language model when generating a new token. The candidate with the lowest perplexity is selected, indicating the highest confidence from the language model. 
These three heuristics apply to 39\%, 56\%, and 5\% of the 79K software entities in \textsc{WikiSER} respectively.

\begin{figure}[t]
    \centering
    \includegraphics[width=0.78\columnwidth]{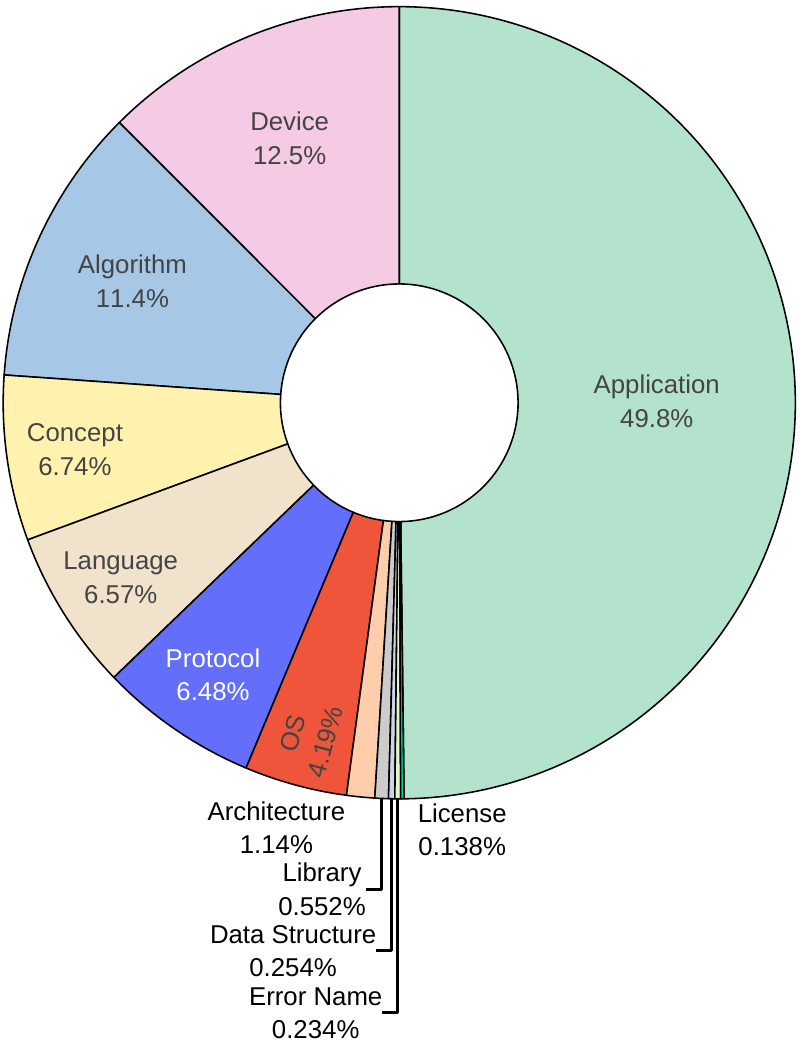}
    \caption{Distribution of Software Entities by Type}
    \label{fig_label_distribution}
\end{figure}

Figure~\ref{fig_label_distribution} shows the distribution of different types of software entities in our dataset. {\em Application} is the most frequently occurring entity type on Wikipedia, comprising 41\% of all software entities present in our dataset. The following are {\em Device},  {\em Algorithm}, {\em General Concept},  {\em Language} and {\em Protocol}, respectively accounting for 12.5\%, 11.4\%, 6.74\%, 6.57\%, and 6.48\% of all entities. 

\subsection{Manual Validation}
\label{sec:manual_validation}

To evaluate the labeling accuracy of our method, we manually validated a random sample from the 1,663,431 sentences identified in Section~\ref{sec:article-collection}. While selecting a sample that is too large is expensive and time-consuming, selecting a sample that is too small can lead to inaccurate conclusions. We use two sampling statistics---confidence level and margin of error---to decide on the proper sample size. We choose a 95\% confidence level and a 5\% margin of error, which are commonly used in empirical software engineering research~\cite{kitchenham2017robust}. This leads to a sample of 387 sentences and 807 corresponding entity labels in the IOB format.

To validate the correctness of these labels, three co-authors hold a 1-hour discussion to establish the span detection and entity categorization criteria and practice on 20 sentences collaboratively to ensure their mutual agreement on the criteria. Then, each manually labels one-third of the 387 sentences. For the tokens that they feel uncertain about, they mark them as ``uncertain'' for later discussion. Then, through cross-validation, they eventually reach a consensus and complete the manual annotation of all sentences. For ``uncertain'' sentences, the three authors conduct a joint discussion to arrive at a final determination for these labels.

By comparing the manual annotations and the auto-generated labels, we find that 74 auto-generated labels (9.17\%) are incorrect. Though this error rate is higher than the 5.38\% error rate of CoNLL~\cite{sang2003introduction}, the most widely used NER in the general text domain, this is reasonable given \textsc{WikiSER}'s fine-grained nature. CoNLL only contains 4 general entity types, while \textsc{WikiSER} includes 12 granular, domain-specific software entity types. The increased specificity makes entity disambiguation more challenging. Furthermore, software entities have high name overlap and aliasing, presenting additional difficulty. Considering \textsc{WikiSER}'s more complex fine-grained distinctions, the labeling quality achieved is acceptable, especially given no existing fine-grained software NER datasets to compare against.

Likewise, we manually validate two notable SER datasets, S-NER~\cite{ye2016software} and SoftNER~\cite{tabassum2020code}, and compute their labeling error rates. For each dataset, we randomly sample 387 sentences, which ensures a 95\% confidence level and a 5\% margin of error, and follow the same procedure to manually label them. Our analysis reveals that the S-NER and SoftNER have 13.93\% and 17.79\% error rates, respectively, as shown in Table ~\ref{tab_comparedata}. Thus, compared with S-NER and SoftNER, our new \textsc{WikiSER} dataset not only has the lowest error rate of 9.17\% but also includes a comprehensive set of software entities and the most labeled sentences (1.7M), compared to 1,015 in S-NER and 7,438 in SoftNER. Though SoftNER has more entity types than \textsc{WikiSER}, 8 of the 20 types in SoftNER are code-related entity types, e.g., {\em class}, {\em variable}, {\em inline-code}, {\em function}, etc. Code-related entities are easier to detect compared with other types of software entities, such as library names and protocols, which have more aliases and naming ambiguity. There is also a large body of literature on recognizing code-related entities~\cite{antoniol2002recovering, bacchelli2010linking, dagenais2012recovering, rigby2013discovering, subramanian2014live, treude2016augmenting, ma2019easy, huo2022arclin}. State-of-the-art techniques such as ARCLIN~\cite{huo2022arclin} have achieved high accuracy in detecting code-related entities. Thus, code-related entities are not of interest in this work.

\begin{table}
\caption{Comparison between \textsc{WikiSER} and two SER Datasets from Stack Overflow}
\centering
\begin{tabular}{@{}llll@{}}
\toprule
 & Ye et al. \cite{ye2016software} & Tabassum et al. \cite{tabassum2020code} & \textsc{WikiSER} \\ 
 \midrule
Entity types & 5 & \textbf{20} & 12 \\
Sentences & 4,646 & 6,510 & \textbf{1.7M} \\
Unique entities & 1,015 & 7,438 & \textbf{79,899} \\
Labeling errors & 13.93\% & 17.79\% & \textbf{9.17\%}\\
\bottomrule
\end{tabular}
\label{tab_comparedata}
\end{table}

\section{Noise-Robust Learning}
\begin{figure*}[t]
    \centering
    \includegraphics[width=0.85\linewidth]{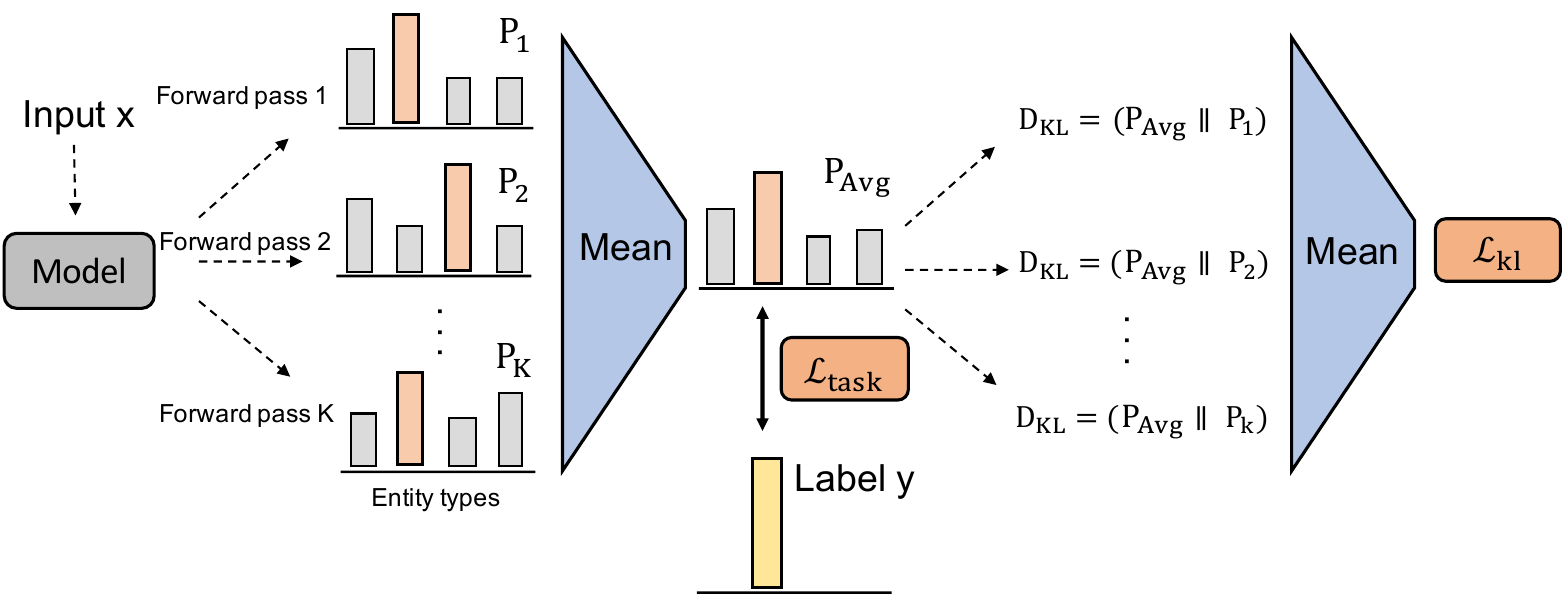}
    \caption{Overview of Self-regularization Framework}
    \label{fig_selfregdiagram}
\end{figure*}

Although \textsc{WikiSER} has a lower labeling error rate compared with other benchmarks, it is not free of labeling errors. Thus, we propose a noise-robust learning framework to account for such labeling errors during model training. The key insight is that, compared to clean-label settings, noisy-label settings can benefit from a ``delayed'' learning curve \cite{zhou2021learning}. This is due to the fact that neural models tend to learn quickly from clean instances that are more compatible with the task's inductive bias at the early stages of training. While doing so, they can become overly confident and less likely to learn from noisier instances that might diverge from the task's inductive bias in later epochs \cite{arpit2017closer}. We can solve this problem by adding a loss term that discourages premature convergence and prevents the model from overfitting uncertain, noisy labels.

Based on this insight, we propose \textit{self-regularization}, a noisy learning approach that leverages the dropout function to reduce overfitting. In deep learning, dropout \cite{srivastava2014dropout} has long been used to improve generalization in neural networks by randomly dropping parts of a network layer. Different versions of the same model induced by the dropout might make different prediction distributions, especially in the presence of noisy instances. We can control this randomness by regularizing the model over its prediction divergence. In \textsc{R-Drop}, Liang et al.~\cite{liang2021rdrop} was the first to use dropout as the main mechanism for regularization in noisy-label settings.  Self-regularization differs from \textsc{R-Drop} in that it relaxes the number of forward passes through the model to be \textit{multiple} instead of just two.

Recall that our approach adds a loss term of self-regularization to the model learning objective. This function accounts for the prediction inconsistency of the model's forward passes during training. 
We explain more details of our training framework below.

\begin{algorithm}[t]
\raggedright
{\textbf{Input:} Train set $\mathcal D=\{(x_i, y_i)\}^N_{i=1}$, agreement loss weight $\alpha$ (hyperparameter), number of forward passes $K$ (hyperparameter).}

{\textbf{Output:} Trained model $M$.}
\begin{algorithmic}[1]
\State Initialize model $M$ from a pretrained language model
\State Warm-up $M$ for $10\%$ of training steps on $\mathcal L_{task}$
\While {M not converged}
    \State Randomly sample a batch $\mathcal B \in \mathcal D$ 
    \State Forward pass $K$ times through $M$ to obtain probability distributions over label space  $\mathcal P = \{P_{j}\}_{j=1}^K$
    \State Compute task loss $\mathcal L_{task}$ by Equation~\ref{eq_taskloss}
    \State Compute KL-divergence loss $\mathcal L_{kl}$ by Equation~\ref{eq_klloss}
    \State Compute $\mathcal L_{agree}$ by Equation~\ref{eq_agreeloss}
    \State Update parameters in $M$ by minimizing $\mathcal L_{agree}$
\EndWhile
\end{algorithmic}

\caption{Training a SER model with self-regularization }
\label{alg_selfregularization}
\end{algorithm}

Let $\mathcal D = \{(x_i, y_i)\}_{i=1}^N$ be a dataset with pairs of an input sequence $x_i$ and a label sequence $y_i$. In the NER setting, an instance in $(x_i, y_i)$ could be considered mislabeled if a token in $x_i$ is wrongly typed (ie. iPhone as \emph{Algorithm} instead of \emph{Device}), or incorrectly assigned a non-named entity (``O''). The goal is to learn a noise-robust model $M$ that tolerates the inevitable training noise.

We initialize $M$ from a pretrained language model such as BERT$ _{base}$, where dropout is by default incorporated. At each training step, we sample a batch  $\mathcal B = \{(x_i, y_i)\}_{i=1}^{|\mathcal B|}$ in $\mathcal D$ for inference on $M$. After each random dropout, $M$ becomes a new submodel with a fraction of the original units in the network. 
The same instance input $x_i$ can outputs different results when passing through $M$. The left-most block in Figure~\ref{fig_selfregdiagram} illustrates this process.

Training of the NER model with self-regularization optimizes two objectives.
The first objective is the divergence loss $\mathcal L_{kl}$. Given the dropout randomness, we obtain a set of $\mathcal P = \{P_{j}\}_{j=1}^K$ distributions over the label space when inputting an instance to $M$ over $K$ forward passes. In noisy settings, $\mathcal P$ is likely to have high variance. We can control for such variance by taking the bidirectional Kullback-Leibler (KL) divergence between the average target probability distribution of $\mathcal P$ and each $P_j$:

\begin{equation}
    \label{eq_klloss}
    \mathcal L_{kl} = \frac{1}{K}\sum^K_{j=1} D_{KL} (P_j || \frac{1}{K}\sum^K_{j=1}P_j)
\end{equation}

\noindent where $\frac{1}{K}\sum^K_{j=1}P_j$ is the average of $K$ probability distributions obtained from the Softmax, denoted as $P_{Avg}$ in Figure~\ref{fig_selfregdiagram}.

The second objective optimizes the cross-entropy task loss $\mathcal L_T$ for $x_i$ for doing NER label classification:
\begin{equation}
    \label{eq_taskloss}
    \mathcal L_{task} = -\frac{1}{K} \sum_{j=1}^{K} \sum_{l=1}^{|x_i|} y_{i, l}\log P_{j, l}
\end{equation}

\noindent where $y_{i,l}$ is the true label of the $l$-th token in input $x_i$ and $P_{j,l}$ is the probability distribution over the label space for the $l$-th token obtained from the $j$-th forward pass. We achieve the standard task loss by averaging the cross-entropy loss over all forward passes.

Finally, the combined agreement loss accounts for both Equation~\ref{eq_taskloss} and Equation~\ref{eq_klloss} as the single learning objective to optimize $M$:

\begin{equation}
    \label{eq_agreeloss}
    \mathcal L_{agree} = \mathcal L_{task} + \alpha \times \mathcal L_{kl}
\end{equation}

\noindent where $\alpha$ is a positive multiplier used to weight the agreement loss. $\mathcal L_{agree}$ is high when the feels uncertain about its prediction and gets smaller when the output probability distribution is more consistent. Algorithm~\ref{alg_selfregularization} describes the full pipeline of self-regularization.

Self-regularization has a few benefits over previous methods. First, compared to existing noisy-label learning methods such as co-regularization and CrossWeigh,\cite{zhou2021learning,wang2019crossweigh}, self-regularization only requires training a single model instead of multiple. This translates to less training time and memory overhead. Second, compared to previous NER models in the software domain, our approach demands neither the use of an external gazette \cite{ye2016software}, nor auxiliary models \cite{tabassum2020code}. Last, the method is versatile in that it can synergize well with any pretrained or randomly initialized models.

\section{Experiments}
\label{sec:experiments}
We design multiple experiments to evaluate our dataset and noisy label learning method. We aim to answer the following research questions:

\begin{itemize}[leftmargin=1.5em]
  \item RQ1: How well does self-regularization work as a denoising measure on \textsc{WikiSER}?
  \item RQ2: How does self-regularization generalize to Stack Overflow benchmarks?
  \item RQ3: What entity types are most difficult to learn?
  \item RQ4: How does the number of forward passes impact self-regularization?
  \item RQ5: How efficient is self-regularization compared to co-regularization?
\end{itemize}

\subsection{Experimental Setup}

\noindent\textbf{Dataset.}
Given the massive size of the \textsc{WikiSER} dataset, we create a subset of it - \textsc{WikiSER$_{small}$} - to train and test SER models. This set strives for a uniform distribution of each entity type. The resulting \textsc{WikiSER$_{small}$} consists of 50K sentences for {\em training}, 8K for {\em validation}, and 8k for {\em test}. 

\vspace{1mm}

\noindent\textbf{Baselines.}
We describe the following baselines to compare with our noisy label learning method.

\begin{enumerate}[leftmargin=1.5em]

    \item \textbf{SoftNER.} SoftNER is the state-of-the-art SER model proposed in \cite{tabassum2020code}. It uses an architecture that combines three embedding attention layers: BERTOverflow, an auxiliary code classifier, and an auxiliary segmentation model to identify name spans. The segmentation model uses extra information such as HTML tags in a SO post, which does not apply to our Wikipedia data. We maintain the code classifier and segmentation model as given, but instead finetuning the entire model on \textsc{WikiSER$_{small}$}.

    \item \textbf{Co-regularization.} Zhou and Chen \cite{zhou2021learning} propose a co-regularization framework that reduces overfitting by regularizing the output divergence of many models, outperforming many methods in information extraction for the general domain. We finetune BERT$_{base}$ with their co-regularization denoising objective as a comparison baseline for self-regularization.

    \item \textbf{BERT$_{base}$.} We finetune pretrained BERT$_{base}$ cased version on our \textsc{WikiSER$_{small}$}. BERT$_{base}$ commonly serves as the standard baseline for many downstream language tasks in the general domain \cite{zhang2022benchmarking, zhou2021learning}.
    
    \item \textbf{RoBERTa$_{base}$.} RoBERTa \cite{liu2019roberta} is another Transformer-based language model that is pretrained from large-scale text corpora. It improves over BERT on many benchmarks by having more diverse training data and modified architecture. We finetune pretrained RoBERTa on  \textsc{WikiSER$_{small}$}.
    
    \item \textbf{BERTOverflow.} Initialized from BERT$_{base}$, BERTOverflow is trained on an additional 152M sentences from Stack Overflow~\cite{tabassum2020code}. In contrast to general-purpose pretrained models, BERTOverflow serves is in-domain for software engineering. We finetune its checkpoint from \cite{tabassum2020code} on \textsc{WikiSER$_{small}$}.

    \item \textbf{Larger model.} BERT$_{large}$~\cite{devlin2018bert} is a bigger variant of BERT$_{base}$ with 340M parameters, whereas the base model has 110M parameters. As the models discussed thus far share the same size, we include a finetuned BERT$_{large}$ baseline to demonstrate the performance of a bigger model on \textsc{WikiSER} and the gains from self-regularization.
\end{enumerate}

\vspace{1mm}
\noindent\textbf{Training details.}
All models use Adam optimizer, learning rate $1e-5$, batch size $16$, dropout rate of $10\%$, and are trained on the NVIDIA RTX A6000 for 30 epochs. For self-regularization, we choose a warm-up rate of $10\%$ and $\alpha=10$.\footnote{We tune $\alpha$ over $\{10, 30, 50\}$ on \textsc{WikiSER$_{small}$} and find $\alpha=10$ to work best. Thus, we use $\alpha=10$ for all models where self-regularization and co-regularization apply.}

\subsection{RQ1: SER Accuracy on \textsc{WikiSER}}
\begin{table}[t]
    \caption{Evaluation results on \textsc{WikiSER}}
    \centering
    \begin{tabular}{llll}
        \toprule
        & P & R & F1 \\
        \midrule
        SoftNER \cite{tabassum2020code} & 64.4 & 69.1 & 66.6 \\
        BERTOverflow \cite{tabassum2020code} & 66.4 & 68.5 & 67.4 \\
        RoBERTa$_{base}$  & 68.2 & 71.0 & 69.6 \\
        BERT$_{base}$ & 68.1 & \textbf{73.1} & 70.5 \\
        BERT$_{base}$ + Co-reg. \cite{zhou2021learning} & 72.7 & 69.1 & 70.8 \\
        BERT$_{base}$ + Self-reg. & \textbf{74.9} & 72.0 & \textbf{73.7} \\
        \midrule\midrule
        BERT$_{large}$ & 69.8 & \textbf{74.5} & 72.1 \\
        BERT$_{large}$ + Self-reg. & \textbf{73.3} & \textbf{74.5} & \textbf{73.9} \\
        \bottomrule
    \end{tabular}
    \label{tab_baselinewiki}
\end{table}

Table~\ref{tab_baselinewiki} shows the results of models trained by our self-regularization framework in comparison to baseline models. Overall, models trained with self-regularization outperform all baselines. This includes BERT$_{base}$ trained with co-regularization by $2.9\%$. The efficacy of both denoising method suggests that we are able to reduce overfitting while learning on \textsc{WikiSER$_{small}$}, which comes with a certain level of noise.
We highlight that BERT$_{base}$ model trained with our self-regularization framework outperforms SoftNER, the SOTA SER model~\cite{tabassum2020code} by 7.1\% in F1 score. SoftNER also performs worst than its pretrained model in BERTOverflow. A plausible reason is that, despite an adaptation to \textsc{WikiSER$_{small}$} via finetuning, SoftNER's auxiliary models might have provided irrelevant signals on clean texts from Wikipedia.

We provide more in-depth analyses of the results.

\vspace{1mm}
\noindent \textbf{Impact of pretrained models.} 

Many NLP studies point to the importance of having pretraining data more closely aligned to the distribution of the downstream task, allowing the model to adapt more easily to the target domain.
For SE, BERTOverflow is a pretrained language model finetuned from BERT$_{base}$ on 152M Stack Overflow posts. Surprisingly, Table~\ref{tab_baselinewiki} shows that BERTOverflow performs the worst, while BERT$_{base}$ performs the best. Notably, BERT$_{base}$ is trained on many general-domain corpora \cite{devlin2018bert} that also includes Wikipedia, the same source in which we construct \textsc{WikiSER}. We suspect that BERT$_{based}$'s good performance can be attributed to the fact that its underlying distribution is closer to \textsc{WikiSER} than that of BERTOverflow. However, BERTOverflow is still effective as a base model for Stack Overflow data, as our experiments demonstrate in Section~\ref{sec:so-experiment}. 

\vspace{1mm}
\noindent\textbf{Impact of model size.} Compared to BERT$_{base}$ (110M parameters) and RoBERTa$_{base}$ (125M), BERT$_{large}$ (340M) trained with self-regularization sees an improvement over its vanilla counterpart. This suggests that gains are possible when the model size increases, though self-regularization shows more effectiveness for smaller models. 
Specifically, in comparison with gains from BERT$_{large}$, BERT$_{base}$ with self-regularization sees an improvement of $3.2\%$ (instead of $1.8\%$) in F1.

\subsection{RQ2: SER Accuracy on Stack Overflow}
\label{sec:so-experiment}

Since Wikipedia imposes strict guidelines and standards for editing and verifying written content, \textsc{WikiSER} tends to not suffer from data noise in the form of spelling mistakes, naming conventions, and others \cite{ye2016software}. In this section, we investigate how our approach generalizes to more noisy benchmarks such as SoftNER~\cite{tabassum2020code} and S-NER \cite{ye2016software}. 

\vspace{1mm}
\noindent\textbf{Datasets.} We use two different Stack Overflow corpora annotated by SoftNER and S-NER. As explained at the end of Section~\ref{section_dataset}, SoftNER has 8 code-related entity types, which is not of interest in this work. Thus, we do not consider them in this experiment. For the remaining 12 entity types, we map them to our entity types for consistency. Please see Supplementary Materials for the full mapping. This results in 9 final entity types, since 3 types are very fine-grained and are merged with other types, e.g., {\em website} merging into {\em application}. For S-NER, which has just 5 entity types (\emph{API}, \emph{Language}, \emph{Platform}, \emph{Framework}, and \emph{Software Standard}), we keep the dataset as it comes, and randomly sample $15\%$ of its data as the test set. We finetune all baseline models separately on these two datasets.

\vspace{1mm}
\noindent\textbf{Results.} Table~\ref{tab_baselinecode} shows that the positive gains from self-regularization also apply to SoftNER and S-NER. BERTOverflow trained with self-regularization performs the best on both datasets. Specifically, it outperforms the self-regularized BERT$_{base}$ model by $11.3\%$ on SoftNER-9 and $3.8\%$ on S-NER in F1. This result makes sense since BERTOverflow is fine-tuned on 152M SO posts and its distribution is more aligned with SoftNER and S-NER than BERT$_{base}$. This implies that in-domain pretraining is still helpful. Additionally, Table~\ref{tab_baselinecode} also suggests that SoftNER performs worse than vanilla BERTOverflow. Since SoftNER adopts auxiliary models to recognize code entities, this implies that these auxiliary models are not as helpful in the absence of code-related entities. 

\begin{table}[t]
    \caption{Evaluation results on two Stack Overflow datasets}
    \centering
    \setlength{\tabcolsep}{5pt}
    \begin{tabular}{l|ccc|ccc}
    \toprule
    \multirow{2}{*}{~} &\multicolumn{3}{c|}{\textbf{SoftNER-9~\cite{tabassum2020code}}} &
    \multicolumn{3}{c}{\textbf{S-NER~\cite{ye2016software}}} \\
     & P & R & F1 & P & R & F1 \\
    \midrule
    BERT$_{base}$ & 64.7 & 64.2 & 64.4 & 77.0 & 80.9 & 78.9 \\
    BERT$_{base}$+Self-reg. & 65.2 & 62.4 & 64.8 & 81.8 & 81.1 & 81.4 \\
    SoftNER & 74.6 & 72.9 & 73.7 & 81.3 & \textbf{84.6} & 82.9 \\
    BERTOverflow & 65.2 & 73.1 & 74.0 & 84.3 & 83.9 & 84.1 \\
    BERTOverflow+Self-reg. & \textbf{75.8} & \textbf{76.5} & \textbf{76.1} & \textbf{86.0} & 84.3 & \textbf{85.2} \\
    \bottomrule
    \end{tabular}%
    \label{tab_baselinecode}
\end{table}

\subsection{RQ3: SER Accuracy by Entity Type}

Table~\ref{tab_baselinebylabel} shows the results of BERT$_{base}$ trained with self-regularization across all entity types. Overall, the model performs fairly well on {\em License}, {\em Error Name}, {\em Data Structure}, {\em Library}, and {\em Operating Systems}. Other entity types, such as {\em Application}, {\em Algorithm}, and {\em General Concept}, appear to be more challenging than others. One possible reason is that entities in those types share more common words with non-SE words, making them more ambiguous to recognize. In contrast, {\em License} and {\em Error Name} tend to be more unique and standardized, making them easier to detect.

\begin{table}[t]
    \caption{Results by entity type. Second column shows the number of entity label occurrences in the test set of \textsc{WikiSER$_{small}$}}
    \centering
    \begin{tabular}{lllll}
        \toprule
        & \# Spans & P & R & F1 \\ 
        \midrule
        General Concept & 2,456 & 67.0 & 62.2 & 64.5 \\ 
        Algorithm & 2,018 & 67.7 & 66.2 & 66.9 \\ 
        Application & 6,861 & 67.9 & 69.7 & 68.7 \\ 
        Device & 3,299 & 73.6 & 69.3 & 71.4 \\ 
        Language & 2,525 & 73.2 & 74.4 & 73.8 \\ 
        Protocol & 2,629 & 74.5 & 73.5 & 74.0 \\ 
        Architecture & 1,538 & 78.0 & 73.8 & 75.8 \\ 
        Operating System & 2,765 & 80.1 & 78.6 & 79.3 \\ 
        Library & 991 & 81.2 & 84.8 & 82.9 \\ 
        Data Structure & 1,051 & 83.1 & 87.4 & 85.2 \\ 
        Error Name & 1,088 & 86.0 & 90.8 & 88.3 \\ 
        License & 1,140 & 86.6 & 90.9 & 88.7 \\
        \midrule
        Micro Avg. & - & 73.8 & 73.5 & 73.7 \\ 
        Macro Avg. & - & 76.6 & 76.8 & 76.6 \\ 
        \bottomrule
    \end{tabular}
    \label{tab_baselinebylabel}
\end{table}

\subsection{RQ4: Choice of $K$ for Self-regularization}
\begin{table}[t]
\caption{Impact of the number of forward passes}
\centering
\begin{tabular}{@{}lccc@{}}
\toprule
\# Forward Passes ($K$) & 2 & 3 & 4 \\ \midrule
BERT$_{base}$ + Self reg. & 70.7 & 73.7 & 73.8 \\
BERT$_{large}$ + Self reg. & 73.9 & 74.0 & 73.8 \\ 
\bottomrule
\end{tabular}%
\label{tab_forwardpasses}
\end{table}
How does increasing the number of forward passes in self-regularization affect model performance? Results from Table~\ref{tab_forwardpasses} show a major improvement for BERT$_{base}$ when the model regularizes over $K=3$ outputs instead of $K=2$. At $K=4$, BERT$_{base}$ sees a marginal improvement of $0.1\%$. For BERT$_{large}$, model performance does not vary greatly with different values of $K$ (within $0.2\%$). Here, we note a potential tradeoff between performance and computational resources. The choice of forward passes can vary by the task and model architecture, and a high $K$ does not necessarily lead to better results compared to smaller values of $K$.

\subsection{RQ5: Training Time and GPU Memory Usage}
\begin{table}
\caption{Training Time and GPU Memory Usage}
\centering
\setlength{\tabcolsep}{2pt}
    \centering
    \begin{tabular}{lrr}
    \toprule
        ~& Wall-clock time (s) & GPU (MB) \\
        \midrule
        BERT$_{base}$ & 287 & 1697 \\
        BERT$_{base}$ + Self-reg. (K=2) & 433 & 1712 \\
        BERT$_{base}$ + Self-reg. (K=3) & 578 & 1718 \\
        BERT$_{base}$ + Co-reg. (N=2) & 512 & 3335 \\
        BERT$_{base}$ + Co-reg. (N=3) & 732 & 5029 \\
        \midrule\midrule
        BERT$_{large}$ & 756 & 5079 \\
        BERT$_{large}$ + Self-reg (K=2) & 1155 & 5079 \\
        \bottomrule
    \end{tabular}
\label{tab_efficiency}
\end{table}

Table~\ref{tab_efficiency} shows the training time and GPU memory usage of models trained with self-regularization compared to co-regularization~\cite{zhou2021learning}. Results show that self-regularization incurs minimal memory overhead (almost at zero additional cost), whereas co-regularization requires 2x the GPU memory when trained with two models and 3x when trained with three models. In wall-clock, self-regularization is also favorable co-regularization while achieving higher F1. It is important to note that the Test F1 on WikiSER for BERT$_base$ + Self-reg outperforms Vanilla BERT$_{large}$, suggesting that the former approach is both better and cheaper. When GPU memory is a concern, self-regularization can be cheaply applied to improve model robustness in noisy-label settings.

\section{Discussion}
\subsection{Threats to Validity}

We discuss the validity of our approach in both data construction and denoising method. First, labeling software entities in sentences could be subjective. Despite our efforts to narrow down a Wikipedia tree for the software engineering domain, we recognize that our annotation still cannot guarantee perfect precision or recall of all relevant named entities from Wikipedia. We estimate the precision of our \textsc{WikiSER} in Section~\ref{section_dataset} and compare our dataset against previous work, which shows that \textsc{WikiSER}'s size and comparatively low label error rate position it as a beneficial contribution for both the software engineering and NLP community. In addition, we understand the limitations of the Wikipedia taxonomy as a rich but not exhaustive source of relevant software entities. Besides Wikipedia, data sources such as GitHub and source code documentation could provide fruitful information.

In model training, we use mostly the same hyperparameters for baseline methods without individual tuning. A more thorough grid search could improve the results for some models in our evaluation. However, we note that while our main method experiments with different $\alpha$, an important hyperparameter for self-regularization, we minimally tune other hyperparameters.

The task of software entity recognition poses many challenges in entity confusion and ambiguity, noisy user-input texts, and constant distribution shifts~\cite{ye2016software}. While we demonstrate the efficacy of the self-regularization framework in recognizing entities for clean (Wikipedia) and noisier user-input texts (Stack Overflow), it is difficult to guarantee that our approach would generalize well to \textit{any} domain. However, we highlight that our framework can be easily adapted to new domains and any pretrained language models without requiring heavy supervision from auxiliary resources.

\subsection{Limitations \& Future Work}
{We evaluate our proposed method on \textsc{WikiSER$_{small}$} rather than the entire WikiSER dataset, which is too large to train and evaluate in our GPU server. Future work could look into training and evaluating SER models with a larger sample from WikiSER or even the entire dataset on more GPUs.

Future work can also look into ways to improve upon more domain-specific methods for noisy label learning and further leverage the massive source of labeled data from \textsc{WikiSER}. The fact that our corpus contains 1.7M sentences makes it an attractive resource for exploring language model pretraining and multi-task learning \cite{aghajanyan2021muppet}. Furthermore, there are many downstream software engineering tasks that could benefit from our noise-robust learning methods for SER, such as traceability link recovery~\cite{antoniol2002recovering, marcus2003recovering, bacchelli2010linking, dagenais2012recovering}, automated documentation~\cite{subramanian2014live, treude2016augmenting, chen2016mining, chen2016techland, li2018improving}, API recommendation~\cite{huang2018api, rahman2016rack, xie2020api}, and bug fixing~\cite{chen2015crowd, gao2015fixing, mahajan2020recommending, mahajan2022providing}. It is worthwhile to augment existing solutions in these downstream tasks with our SER model. Finally, given the recent advancement in large language models (LLMs) such as ChatGPT, it is interesting to investigate how well LLMs perform in software entity recognition tasks.

\section{Conclusion}
In this work, we construct \textsc{WikiSER}, a large and high-quality software entity recognition dataset by leveraging the Wikipedia corpus. To account for labeling errors in SER datasets, we propose a new noise-robust learning method called self-regularization. Compared with multiple baseline models, including a SOTA SER model~\cite{tabassum2020code} and a SOTA noise-robust learning method~\cite{zhou2021learning}, models trained with self-regularization perform the best while being more computationally efficient.  Furthermore, self-regularization also generalizes well to two existing SER datasets from Stack Overflow. Finally, we highlight several improvement opportunities and outline future work.

\section*{Acknowledgment}
The authors would like to thank the anonymous reviewers for their valuable comments. This research was in part supported by an Amazon Research Award and a Cisco Research Award.

\bibliographystyle{IEEEtran}
\bibliography{ref}

\end{document}